\documentstyle[festkoer,psfig]{vieweg}

\author{Johann Kroha}

\Kurzautor{J. Kroha}

\title{Energy and phase relaxation in non-equilibrium diffusive
nano-wires with two-level systems}

\Kurztitel{Energy and phase relaxation in nano-wires}

\Adresse{Institut f\"ur Theorie der Kondensierten Materie\\
Universit\"at Karlsruhe, Postfach 6980, 
D-76128 Karlsruhe, Germany}

\begin{document}

\Titel

\begin{abstract}
In recent experiments
the non-equilibrium distribution function $f(E,U)$ in diffusive 
Cu and Au quantum wires at a transport voltage $U$ shows  
scaling behavior, $f(E,U)=f(E/eU)$, indicating a non-Fermi 
liquid interaction with {\it non-vanishing} $T=0$
scattering rate. The two-channel Kondo (2CK) effect, 
possibly produced by degenerate two-level systems, 
is known to exhibit such behavior. Generalizing the auxiliary 
boson method to non-equilibrium, we calculate $f(E,U)$ in the 
presence of 2CK impurities. We show that the 2CK 
equations reproduce the scaling form $f(E/eU)$. For all 
measured samples the theoretical, scaled distribution 
functions coincide quantitatively with the experimental results, 
the impurity concentration being 
the only adjustable parameter. This provides a microscopic 
explanation for the experiments and, considering that no
other mechanism producing the scaling form is known to date, 
lends strong evidence for the presence of degenerate
two-level defects in these systems. 
The relevance of these results for the problem of 
dephasing in mesoscopic wires is discussed. 
\end{abstract}

\vspace*{0.4cm}
\section{Introduction}
Two-level systems (TLS) have been known to exist in disordered
solids for a long time. Their signatures have been observed in
the anomalous thermodynamic properties of glasses, which may be
explained by a flat distribution of level splittings. In metals,
slow two-level fluctuators are evidenced by telegraph noise 
in the conductance \cite{telegraph}. A new physical phenomenon, the two-channel
Kondo (2CK) effect \cite{nozieres.80,coxzawa.98}, arises when a fast, 
energetically (nearly)
degenerate TLS is embedded in a metal, where the local impurity degree
of freedom couples to the continuum of conduction electrons via
an exchange interaction. The channel degree of freedom, conserved
by this interaction, is the magnetic conduction electron spin,
which is always degenerate in the presence of time reversal symmetry.

The 2CK effect exhibits striking non-Fermi liquid behavior 
\cite{andrei.84,wiegmann.83,schlottmann.93} with 
a non-vanish\-ing single-particle scattering rate at the Fermi energy
$\varepsilon_F$ for temperature $T=0$, logarithmic behavior of the 
linear specific heat coefficient and the susceptibility, 
and a non-analytic, universal correction to the electronic density
of states, $\Delta N(E)/(B\sqrt{T}) = h(x)$, where $B$ is a non-universal 
constant, $x:=|E-\varepsilon_F|/T$ and $h(x)$ is a universal scaling 
function with $h(x)=\sqrt{x}$ for $x\gg 1$ \cite{affleck.93,hettler.94}. 
Signatures of the latter have been observed
as zero bias conductance anomalies (ZBA) of nano point contacts
\cite{ralph.92,ralph.94,upad.97}. 
As suggested in Ref. \cite{zawa.99}, the finite 
single-particle scattering rate at the Fermi energy would provide a natural
explanation for the saturation of the dephasing time $\tau _\varphi$
at low temperatures,
which has recently been observed \cite{mohanty.97} in magnetotransport 
measurements of weak localization in disordered wires.

However, the actual existence of the 2CK effect in nature has remained
a controversial issue, partly because the ZBA could also be explained
qualitatively \cite{wingreen.95}, although not quantitatively
\cite{hettler.94,upad.97,ralph.95,altshuler.98}, by the 
Al'tshuler-Aronov diffusion-enhanced Coulomb interaction \cite{altaronov.79}, 
partly because the physical realization of degenerate TLS is poorly
understood. Therefore, it is essential to develop unambiguous 
methods for detecting 2CK physics in mesoscopic systems.

Recently, it has been demonstrated in a landmark experiment performed
by the Saclay ``Quantronics'' group \cite{pothier.97}
that unique information about the electronic interactions
in a metal may be extracted from the distribution function
$f(E,U)$ of quasiparticles with energy $E$ when the system is driven
far away from equilibrium by a transport voltage $U$: The shape of
$f(E,U)$ is determined by the energy dependence of relaxation processes
which tend to equilibrate the system. It has been 
found that in diffusive Cu and Au 
nano-wires the distribution function obeys a scaling form,
$f(E,U)=f(E/eU)$, implying a non-Fermi liquid interaction
\cite{pothier.97,pothier.99}. 

In this article we discuss the close relation between this 
peculiar scaling property, which we will call
``$E/eU$ scaling'' for brevity, 
and non-Fermi liquid behavior. It is then shown that 
the 2CK effect obeys $E/eU$ scaling and reproduces quantitatively the
measured distribution functions, while any other type of interaction
is ruled out. This provides the strongest case to date for the 
physical realization of the 2CK effect in Cu and Au nano-wires
induced by TLS. We briefly discuss its relation \cite{gougam.00}
to the problem of the dephasing time saturation at low temperatures
\cite{mohanty.97}.

\section{Experiment and non-Fermi liquid signature}

\begin{figure}
\vspace*{0.7cm}
\hspace*{1.4cm}\psfig{figure=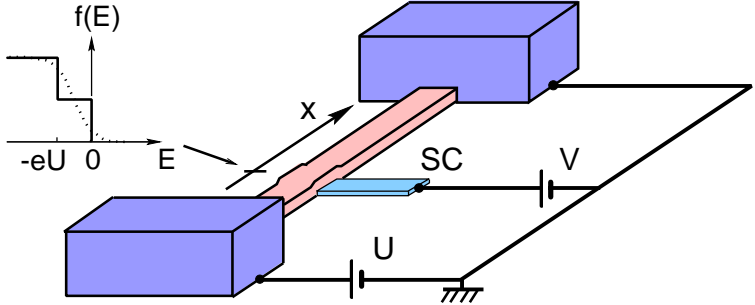,width=10.0cm}\hfill
\caption[fig1]{Experimental setup for measuring the non-equilibrium 
distribution function in a metallic nano-wire.}
\label{expsetup}
\end{figure}

The distribution function $f(E,U)$ was measured \cite{pothier.97}
in an experimental 
setup where a non-equilibrium current was driven through a Cu nano-wire
contacted by two reservoirs at chemical potentials $0$ and $eU$,
respectively (Fig.~\ref{expsetup}). In addition, a superconducting Al
tunneling junction was attached at a position $x$ along the wire,
the Al slab being in equilibrium with itself. For a voltage
$V$ across the junction the tunneling current is given by 
\begin{equation}
j_{tunnel} = \frac{e}{\hbar} |t|^2 \sum _{\sigma}
\int dE \bigl[ f(E) - f^o(E+eV) \bigr]
N_{Cu,\sigma}(E)\; N_{sc}(E+eV)\ ,
\label{jtunnel}
\end{equation}
where $t$ is the (energy independent) tunneling matrix element,
$f^o(E)$ is the Fermi distribution function in the superconductor,
and $N_{Cu}$, $N_{sc}$ denote the density of states in the wire and in 
the superconductor, respectively. Since for the voltages used in the
experiment $N_{Cu}$ is flat and the BCS density of states $N_{sc}$
is measured independently, the non-equilibrium distribution $f(E)$ in the Cu 
wire can be extracted from this expression. The electronic transport in the
wire is diffusive with diffusion coefficient $D$. Length $L$ and thickness $d$
of the wire are such that the Fermi surface is three dimensional
and the Coulomb and phonon scattering times are large compared to the
electronic diffusion time $\tau _D = L^2/D$ through the wire,
so that equilibration due to these processes can be neglected
\cite{pothier.97}.
In this situation, assuming purely elastic scattering, one expects
the distribution function at a given position $x$ along the wire 
to be a linear superposition of the Fermi functions in the reservoirs,
\begin{equation}
f^{elast}_x(E,U) = \bigl(1-\frac{x}{L}\bigr) f^o(E+eU) + \frac{x}{L} f^o(E)\ ,
\label{felast}
\end{equation}
since there is no energy exchange within the electron system or between
the electron system and the lattice. Eq.~(\ref{felast}) is a solution of
the diffusive Boltzmann equation (\ref{boltz}) \cite{nagaev.92}, 
when the collision 
integral vanishes (see section 3).
This situation is to be distinguished from the
hot electron regime, where local equilibration occurs due to inelastic
processes \cite{kozub.95}. 

The measured distribution functions showed rounding of the Fermi steps
as compared to Eq.~(\ref{felast}) and obeyed scale invariance with
respect to the transport voltage $U$,  
$f(E,U)=f(E/eU)$, when $U$ exceeded a certain 
low energy scale, $eU \stackrel{>}{\sim} 0.1meV$ \cite{pothier.97}.
Deviations from scaling were observed again for voltages larger than
a high energy scale $E_o \simeq 0.5meV$. 
The latter may be explained by reservoir 
heating effects or by the electrons coupling to additional degrees of freedom 
at high energies. We now deduce the non-Fermi liquid
signature from the scaling property. The latter implies that the equation of
motion for $f(E)$, the Boltzmann equation and, as a consequence, the inelastic
single-particle collision rate $1/\tau (E)$ are scale invariant.
Assuming a (yet to be determined) two-particle potential $\tilde V(\varepsilon )$
with energy transfer $\varepsilon$,
$1/\tau$ is given in 2nd order perturbation theory as \cite{remark}
\begin{equation}
\frac{1}{\tau (E)} \equiv \frac{1}{\tau (E/eU)} \simeq
N_{Cu}(0)^3 \int d\varepsilon  \int d\varepsilon '
|\tilde V (\varepsilon )|^2 \tilde F \Bigl(
\frac{E}{eU},\frac{\varepsilon}{eU},
\frac{\varepsilon '}{eU} \Bigr) \ . 
\label{perturb}
\end{equation}
Here $\tilde F$ is a combination of distribution functions $f$ guaranteeing
that there is only scattering from an occupied into an unoccupied state.
Therefore, the experimental results about the scaling property of $f$
imply that $\tilde F$ depends only on the
dimensionless energies as displayed in Eq. (\ref{perturb}). 
Demanding scale invariance
for $1/\tau$ with respect to $eU$, i.e. making the frequency integrals
dimensionless, implies a characteristic energy dependence of the 
interaction and the single particle scattering rate,
$\tilde V(\varepsilon ) \propto 1/\varepsilon $ \cite{pothier.97}
and $1/\tau (E) \propto -{\rm ln} (E/E_o)$, for energies less than $E_o$. 
These infrared singularities indicate a breakdown of Fermi liquid theory
within the 2nd order perturbation theory argument applied here.

\section{\hspace*{-0.1cm}Two-channel Kondo effect and scaling in non-equilibrium}
The strong infrared divergence of the two-particle 
potential, $\tilde V (\varepsilon )
\propto 1/\varepsilon $, deduced above, is not explained by any conventional 
interaction, including the Al'tshuler-Aronov interaction
\cite{altaronov.79}. It must, therefore,
be generated by an infinite resummation of logarithmic terms
obtained in perturbation theory due to the presence of a Fermi edge. 
In this section the 2CK effect is briefly reviewed. We then show that the
effective electron-electron vertex, mediated by a 2CK impurity, has a
$1/\varepsilon $ divergence and calculate the resulting 
distribution functions away from equilibrium. 

The 2CK effect arises whenever a local, energetically degenerate
two-level degree of freedom (pseudospin $\tau=\pm 1/2$)
is coupled to a system of two identical
conduction electron bands or channels via a pseudospin exchange interaction,
which, however, conserves the channel degree of freedom. 
As for the physical realization of the pseudospin, it has been 
suggested \cite{coxzawa.98} that it might arise from the
degenerate positions of an atom in a symmetrical double well potential,
from a rotational degree of freedom of a lattice defect or group of atoms,
or from sliding kinks on screw dislocations in a lattice 
\cite{sethna.00}. The channel degree of freedom is then the magnetic
electron spin $\sigma =\pm 1/2$, which is necessarily degenerate because of
time reversal symmetry (Kramers degeneracy).

In order to describe non-equilibrium situations,
it is useful to represent this system in terms of the low-energy
physics of a SU(2)$_{\rm pseudospin}$$\times$SU(2)$_{\rm channel}$
Anderson impurity model in the Kondo limit,
i.e. in terms of a doubly degenerate local level $E_d<0$,
coupled via a hybridization $v$ to two identical conduction channels, 
with conduction electron
operators $c^{\phantom{\dag}}_{k\tau\sigma }$, $c^{\dag}_{k\tau\sigma }$. 
The dynamics of the two-level degree of freedom (pseudospin) 
is described by a fermionic field operator $d^{\dag}_{\tau\sigma}$, with
the restriction 
\begin{figure}
\vspace*{0.7cm}
\hspace*{1.4cm}\psfig{figure=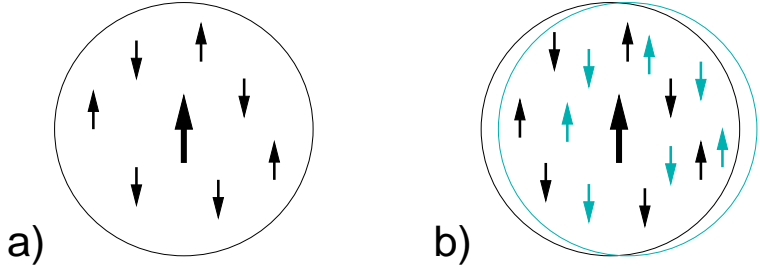,width=10.0cm}\hfill
\caption[fig1]{Schematic snapshot of the Kondo screening cloud.
The bold arrows represent the impurity (pseudo)spin.
a) Single-channel Kondo effect. Conduction electrons and local impurity
spin form a collective singlet ground state with entropy $S(T=0)=0$.
b) Two-channel Kondo effect. Each of the two
identical conduction electron bands form a screening cloud (black and grey),
so that the net pseudospin of the combined clouds is not 0; 
pseudospin singlet formation is
frustrated. The ground state is not unique, resulting in a non-vanishing
zero-point entropy $S(T=0)=k_B {\rm ln} \sqrt{2}$.}
\label{2CKcartoon}
\end{figure}
that the local level must not be doubly occupied at any time.
The latter is implemented exactly by decomposing the local
operator into auxiliary fermion $f^{\dag}_{\tau }$
and boson $b^{\phantom{\dag}}_{\bar\sigma}$ operators,  
$d^{\dag}_{\tau\sigma} = f^{\dag}_{\tau } b^{\phantom{\dag}}_{\bar\sigma}$,
supplemented by the operator constraint
$\hat Q = \sum _{\tau} f^{\dag}_{\tau } f^{\phantom{\dag}}_{\tau }
+\sum _{\sigma} b^{\dag}_{\bar\sigma} b^{\phantom{\dag}}_{\bar\sigma} \equiv 1$
 \cite{barnes.76}. Thus, we describe the 2CK system by the Anderson hamiltonian 
in auxiliary particle representation,
\begin{equation}
H = \sum _{k\sigma\tau} 
\varepsilon _k c^{\dag}_{k\tau\sigma } c^{\phantom{\dag}}_{k\tau\sigma }
+E_d \sum _{\tau} f^{\dag}_{\tau } f^{\phantom{\dag}}_{\tau }
+v \sum _{k\sigma\tau}( f^{\dag}_{\tau } b^{\phantom{\dag}}_{\bar\sigma}
                       c^{\phantom{\dag}}_{k\tau\sigma } +h.c.) \ .
\label{hamiltonian}
\end{equation}
Note that the boson field $b_{\bar\sigma}$ transforms with respect to the
adjoint representation of SU(2)$_{\rm channel}$ (denoted by the index
$\bar\sigma$).
By integrating out $ b^{\phantom{\dag}}_{\bar\sigma}$
in the Kondo regime ($\langle\sum _{\tau}
 f^{\dag}_{\tau } f^{\phantom{\dag}}_{\tau } \rangle \simeq 1$, $|E|\ll |E_d|$),
$H$ is reduced to a 2CK model where the exchange coupling $J=|v|^2/|E_d|$
is mediated by the boson propagator and, as required, conserves the
channel degree of freedom $\sigma$. 
The above auxiliary particle 
representation is known to describe the infrared behavior of the 2CK
effect correctly \cite{cox.93}
already on the level of a leading-order selfconsistent
perturbation theory in the hybridization $v$, the non-crossing
approximation (NCA) \cite{bickers.87}.  
It also allows for the application of standard Green's function methods 
and, in particular, is readily generalized \cite{hettler.94} 
to non-equilibrium by means of the Keldysh technique.

Physically, the presence of two identical
conduction electron channels leads to an overscreening of the impurity 
pseudospin below the Kondo temperature
$T_K = \bigl( 2 \Gamma/\pi \bigr) {\rm exp}
\bigl(- \pi |E_d|/2\Gamma\bigr)$, $\Gamma = \pi |v|^2 N_{Cu}(0)$,
as elucidated in Fig.~\ref{2CKcartoon}. This results in a non-vanishing
zero-point entropy $S(T=0)=k_B{\rm ln}\sqrt{2}$
\cite{andrei.84} and is closely
related to the finite single-particle decay rate $1/\tau (E=0,T=0) >0$
predicted by the 2CK effect.

\begin{figure}
\vspace*{0.7cm}
\hspace*{1.4cm}\psfig{figure=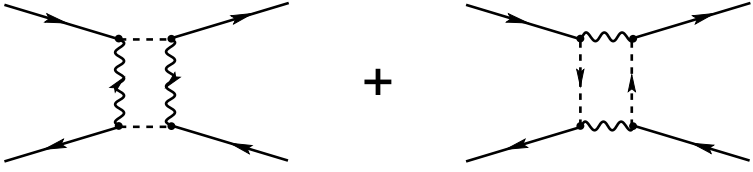,width=10.0cm}\hfill
\caption[fig1]{Leading contributions to  the effective electron-electron
vertex induced by a 2CK impurity. Solid, dashed, and wiggly lines
correspond to conduction electron, pseudofermion, and slave boson
propagators, respectively.}
\label{eff_int}
\end{figure}

Below the Kondo scale $T_K$, 
the auxiliary particle propagators $G_f$, $G_b$ are known to 
exhibit infrared power-law behavior in equilibrium,  
$G_f(\omega ) \propto \omega ^{-\alpha _f}$,
$G_b(\omega ) \propto \omega ^{-\alpha _b}$ for $\omega 
{\stackrel {<}{\sim}}T_K$ \cite{muha.84}, where the exact 
\cite{cox.93,affleck.93} relation 
$\alpha _f + \alpha _b = 1$ is characteristic for the multichannel
Kondo effect ($\alpha _f = M/(M+N)$, $\alpha _b = N/(M+N)$ for pseudospin
degeneracy $N$ and channel number $M$). Using power counting arguments,
it follows that each fermion (boson) propagator contributes a power
$-\alpha _f$ ($-\alpha _b$) and each frequency integral contributes
a power 1 to the frequency dependence of 
the effective electron-electron vertex $\gamma (\omega )$,
shown in Fig.~\ref{eff_int}. Thus, the 2CK electron-electron 
vertex depends on the frequency transfer $\omega $ as
$\gamma (\omega) \propto \omega ^{1-2(\alpha _f + \alpha _b)} = \omega ^{-1}$
\cite{remark2}. On the perturbative level, this implies scale invariance 
of the inelastic scattering rate $1/\tau (E)$ with respect to
the transport voltage $U$, as discussed in section 2. A self-consistent
resummation of these terms in non-equilibrium shows that the scale 
invariance persists up to single-particle energies $E$ of the order of 
$eU$ \cite{kroha.00}. 
It may be shown that the range of eU values for which scaling
behavior is obeyed is bounded from below by the Kondo temperature $T_K$.   
For large bias voltage, deviations from scaling occur when $eU$ becomes
comparable to the band cutoff or to the local level $|E_d|$.
\begin{figure}
\hspace*{1.4cm}\psfig{figure=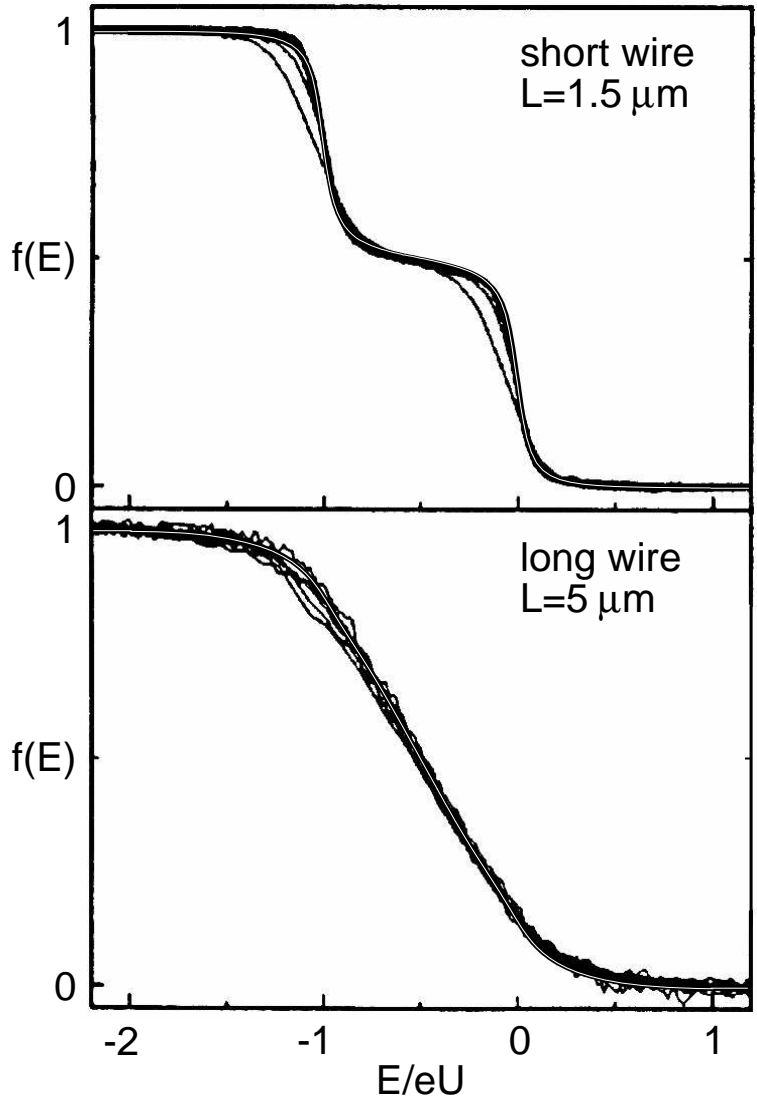,width=10.0cm}\hfill
\caption[fig1]{Scaled non-equilibrium distribution functions for a short
($L=1.5\mu m$, diffusion const. $D=65cm^2/s$) and a long
($L=5.0\mu m$, diffusion const. $D=45cm^2/s$) Cu nano-wire.
Black curves: Experimental data for $U$ in a range  
between 0.05mV and 0.3mV in steps of 0.05mV, 
taken from Ref. \cite{pothier.97}. The two curves showing deviations from
scaling are at small voltages, 0.05mV and 0.1mV, respectively.
Light curves: Theory in the scaling regime, $eU\gg T_K$.
The 2CK impurity concentration $c_{imp}$ obtained from the fit is
approximately $8\cdot 10 ^{-6}$/(lattice unit cell).} 
\label{f_expth_s}
\end{figure}

The remaining task is to calculate the non-equilibrium distribution function
$f_x(E,U)$ at an arbitrary position $x$ along the wire 
(compare Fig. \ref{expsetup}) in the presence of 2CK defects.
In the regime where the inelastic mean free path $\ell _{inel}$,
approximately given by the average spacing $a$ between 2CK impurities,
is small 
\begin{figure}
\hspace*{1.4cm}\psfig{figure=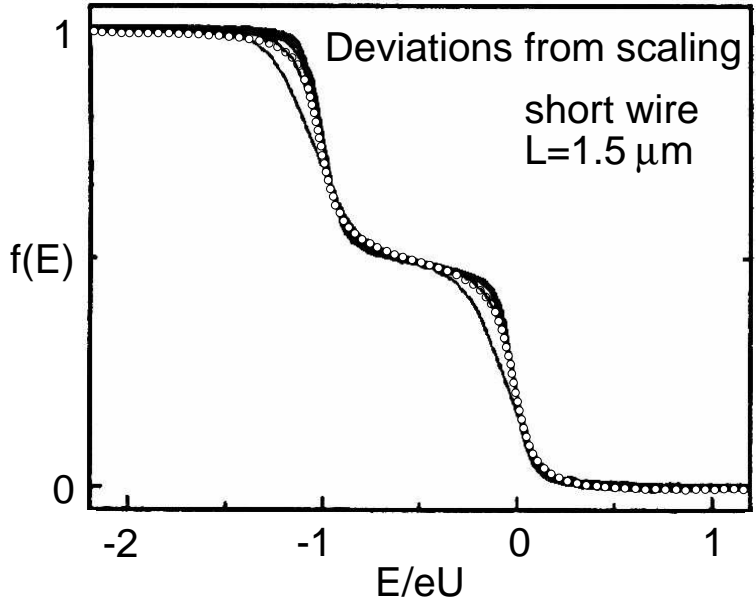,width=10.0cm}\hfill
\caption[fig1]{Deviations from scaling. Experimental data are the
same as in Fig. 4, upper panel. Open circles represent the fit of the
theory to the first experimental distribution function which shows deviations 
from scaling at small voltage, eU=0.1meV.
Since the 2CK impurity density $c_{imp}$ has been determined as the only 
adjustable parameter in the scaling regime (Fig. 4), for given
transport voltage $U$, the Kondo temperature $T_K$ (or the temperature $T$,
whichever is larger), is the only parameter that controls deviations from
scaling at small voltage. It may be detetermined from the fit as
$T_K \simeq 1K$.}
\label{f_expth_ds}
\end{figure}
compared to the wire length, but large compared to the
elastic mean free path $\ell $, $L\gg \ell _{inel} \simeq a \gg \ell$,
the dynamics of a given 2CK impurity is determined by interactions 
with electrons 
probing many different spatial regions in the wire. 
It is, therefore, appropriate to first calculate the averaged quasiparticle
distribution $\overline{f_x(E,U)}$ as the solution of the impurity ensemble
averaged Boltzmann equation, subject to the boundary conditions
$\overline{f_{x=0}(E,U)} = f^o(E+eU)$, $\overline{f_{x=L}(E,U)} = 
f^o(E)$.
The equation of motion for $\overline f$ is obtained by summing
over all local quasiparticle momenta $p$ and exploiting the fact that
the current in the disordered system is diffusive,
\begin{equation}
\vec j _x(E) =\sum _p \frac{p}{m}  \overline{f_{x}(E,U)} = - D \nabla _x 
\rho _x(E) = - D \nabla _x  \sum _p  \overline{f_{x}(E,U)}. 
\end{equation}
Here, $\nabla _x$ denotes the component of the gradient in the
direction along the wire. In the stationary case, 
the resulting diffusive Boltzmann equation reads \cite{nagaev.92} 
\begin{equation}
\nabla _x^2 \overline{f_{x}(E,U)} = -\frac {1}{D} {\cal C}_{2CK}
\bigl[ \overline {f_{x}(E,U)} , c_{imp} \bigr] ,
\label{boltz}
\end{equation}
where ${\cal C}_{2CK}$ is the collision integral due to 2CK scattering. 
For small impurity concentration $c_{imp}$ is proportional to $c_{imp}$.
Since the diffusion constant $D$ is 
measured experimentally, $c_{imp}$
is the only adjustable parameter of the theory in the scaling regime.

In the experiment, the tunneling current Eq.~(\ref{jtunnel}) measures 
the local (unaveraged) distribution function of electrons in the vicinity
of the junction. It is determined by the local 
stationarity condition that it must be equal to the distribution function
of the 2CK impurity states. The results are shown in 
Fig.~\ref{f_expth_s} and display scaling behavior for
$eU\gg T_K$. It is seen that there is excellent
quantitative agreement between theory and experiment for all samples.
Considering the fact that no other interaction producing $E/eU$ scaling  
is known to date, this provides strong evidence for the presence of 
2CK impurities in evaporated Cu nano-wires.
From the fit of the theory to that experimental curve which 
for the first time shows deviations from scaling as the voltage is
decreased, i.e. for $eU=0.1meV$ (Fig. 5), the 
corresponding experimental low energy scale may be identified
with the low energy scale of the theory, the 
Kondo temperature $T_K$. Thus we have $T_K \simeq 1K$, which is in rough
agreement with earlier experimental results on TLS in Cu point contacts
\cite{ralph.92,ralph.94}.

\section{Relation to dephasing}
In equilibrium, the non-vanishing 2CK quasiparticle scattering rate
$1/\tau (E)$ cross-es over to a pure dephasing rate $1/\tau _{\varphi}$,
as the quasiparticle energy  approaches the Fermi surface, $E,T\to 0$,
since at the Fermi energy no energy exchange is possible. 
One might, therefore, conjecture that degenerate TLS could be the
origin of the dephasing time saturation observed in magnetotransport
measurements of weak localization. This assumption is indeed supported by
several coincidences between the dephasing time measurements and the  
results on the non-equilibrium distribution function: (1) The
dephasing time $\tau_{\varphi}$ extracted from magnetotransport
experiments \cite{gougam.00,mohanty.97} 
is strongly material, sample, and preparation dependent.
This suggests a non-universal dephasing mechanism, like TLS, which is
not inherent to the electron gas.
(2) The dephasing time in Au wires is generically shorter 
than in Cu wires \cite{pothier.99}.
This is consistent with the fact that the estimates
for the TLS concentration $c_{imp}$, obtained from the
fit of the present theory to the experimental distribution functions,
is much higher in Au than in Cu wires \cite{kroha.00}.
(3) In Ag wires one observes neither dephasing saturation 
nor $E/eU$ scaling of the distribution function. 
This is consistent with the assumption that there are no 2CK defects
present in the Ag samples \cite{pierre.00}.

\section{Concluding remarks}

We have calculated the quasiparticle distribution as a function of
the excitation energy $E$ in diffusive nano-wires in the presence of
2CK impurities, when the system is driven far away from equilibrium
by a finite transport voltage $U$. The present theory reproduces the
experimental finding that in Cu and Au wires 
the nonequilibrium distribution function
displays the scaling property $f(E,U)=f(E/eU)$ for $eU$ above a low-energy
scale $T_o$. Within the theory, $T_o$ is given by the
Kondo temperature $T_K$ or by $T$, whichever is larger.
The 2CK impurity density is the only adjustable parameter of the theory
in the scaling regime. The quantitative agreement between the present
theory and experiment and the fact that the experimental scaling property
is not even qualitatively explained by any other type of interaction,
provide strong evidence for the existence of 2CK impurities in Cu and Au
wires, while there seem to be no such defects present in Ag wires.
We have not attempted here to give a microscopic model for the
physical realization of 2CK defects. For that purpose, it should be
useful to perform numerical simulations \cite{sethna.00}
of dislocations in the respective materials.\\

\vspace*{-0.75cm}
\section*{Acknowledgements}
It is a pleasure to thank A. Zawadowski, H. Pothier, B. L. Al'tshuler,
D. Esteve,
J. v. Delft, and P. W\"olfle for stimulating and fruitful discussions.
This work is supported by DFG and by grants of 
computing time from the J. von Neumann
Institute for Computing (NIC) J\"ulich and from the HLRZ Stuttgart.


\begin{thebibliography}{99}
\bibitem{remark} It has been verified \cite{pothier.97} that the experimental
results are consistent with a spatially local interaction 
$\tilde V(\varepsilon )$.
\bibitem{remark2} It can be shown that this infrared power-law behavior
is reproduced in arbitrarily high order in the coupling $\Gamma$.
\end{thebibliography}
\end{document}